\documentclass[a4paper,aps,prl,showpacs,twocolumn,superscriptaddress]{revtex4-1}

\usepackage[utf8]{inputenc} 
\usepackage[T1]{fontenc}    
\usepackage[british]{babel}  
\usepackage{lmodern} 
\usepackage[scaled=1.03]{inconsolata}
\usepackage[usenames,dvipsnames]{color}
\usepackage[colorlinks,citecolor=blue,linkcolor=magenta,urlcolor=blue]{hyperref}
\usepackage{graphicx}
\usepackage{tikz}
\usepackage[babel]{microtype}
\usepackage{amsmath,amssymb,amsthm,bm,mathtools,amsfonts,mathrsfs,bbm,dsfont}
\usepackage{xspace}
\usepackage{multirow}
\usepackage{verbatim}
\usepackage{float}

\usepackage{physics}
\newcommand{\id}{\ensuremath{\mathds{1}}}
\usepackage{bbold}
\usepackage{tcolorbox}

\newcommand{\C}{\mathbb{C}}

\newcommand{\mc}[1]{\mathcal{#1}}
\newcommand{\ms}{}

\newcommand{\kb}[2]{|#1\,\rangle\langle\,#2|}
\def\<{\langle}
\def\>{\rangle}

\newcommand{\fii}{\varphi}

\newtheoremstyle{mystyle}
  {6pt}
  {6pt}
  {\normalfont}
  {0pt}
  {\bf}
  {.}
  { }
  {}

\theoremstyle{mystyle}

\newtheorem{observation}{Observation}
\newtheorem{remark}{Remark}

\usepackage{dblfloatfix}
\usepackage{blindtext}

\usepackage{xcolor}
\colorlet{myPurple}{blue!40!red}
\colorlet{myCyan}{cyan!50!gray}

\definecolor{quantumviolet}{HTML}{53257F} 
\definecolor{quantumgray}{HTML}{555555} 
\definecolor{mygray}{gray}{0.95} 

\newtcolorbox[auto counter,number within=section]{boxfigure}[2][]{
colback=mygray,colframe=myPurple,fonttitle=\bfseries,width=\columnwidth,float*=ht,lower separated=false, halign=justify,title=Box~\thetcbcounter: #2,#1}

\begin{document}
\nonfrenchspacing
\title{Retrievability of information in quantum and realistic hidden variable theories}

\author{Roope Uola}
\affiliation{Department of Applied Physics, University of Geneva, 1211 Geneva, Switzerland}

\author{Erkka Haapasalo}
\affiliation{Centre for Quantum Technologies, National University of Singapore, Science Drive 2 Block S15-03-18, Singapore 117543}

\author{Juha-Pekka Pellonp\"a\"a}
\affiliation{Department of Physics and Astronomy,
  University of Turku, FI-20014 Turun yliopisto, Finland}

\author{Tom Kuusela}
\affiliation{Department of Physics and Astronomy,
  University of Turku, FI-20014 Turun yliopisto, Finland}

\date{\today}

\begin{abstract}
We propose a generalisation of the Leggett-Garg conditions for macrorealistic behaviour. Our proposal relies on relaxing the postulate of non-invasive measurability with that of \textit{retrievability of information}. This leads to a strictly broader class of hidden variable theories than those having a macrorealistic description. Crucially, whereas quantum mechanical tests of macrorealism require one to optimise over all possible state updates, for retrievability of information it suffices to use the basic L\"uders state update, which is present in every quantum measurement. We show that in qubit systems the optimal retrieving protocols further relate to the fundamental precision limit of quantum theory given by Busch-Lahti-Werner error-disturbance uncertainty relations. We implement an optimal protocol using a photonic setting, and report an experimental violation of the proposed generalisation of macrorealism.
\end{abstract}

\maketitle

\textit{Introduction.---} The concept of macrorealism aims at describing temporal correlations based on two classical assumptions: \textit{macrorealism per se}, i.e.\ the existence of macroscopic states of the system, and \textit{non-invasive measurability}, i.e.\ the possibility of measuring these states without disturbing them or the subsequent dynamics. Famously, the conjunction of these two assumptions is in contradiction with quantum theory, as witnessed by violations of the celebrated Leggett-Garg inequalities \cite{leggett85,Emary13}.

Practical tests of macrorealism suffer from a so-called clumsiness loophole. In short, the loophole states that violations of macrorealism may be caused by the lack of control over one's measurement apparatuses. In other words, one might intentionally or unintentionally violate the non-invasiveness assumption. As an example, take a sequence of two measurements of
the spin of a quantum particle in a given direction.
If the first measurement disturbs the system by, e.g., introducing a unitary rotation, the outcome of the second measurement may be affected. In such case, one is in clear contradiction with the non-invasiveness assumption.

In order to go around the clumsiness loophole, various different ways of justifying that one’s measurement apparatus is not affecting the subsequent measurement statistics have been introduced. To mention a few, ideal negative-choice measurements were discussed by Leggett and Garg \cite{leggett85}, adroit measurements (resp. $n$-term adroit measurements) and were introduced in Ref.~\cite{wilde11} (resp. Ref.~\cite{uola19a}), non-disturbing measurements in Ref.~\cite{george13}, and auxiliary control measurements were used in Ref.~\cite{knee16}. Recently, there has also been a proposal towards formalising macrorealism in terms of generalized probability theories \cite{Schmid22}, in which violations of macrorealism can be approached through the process of theory-agnostic tomography \cite{Grabowecky22,Mazurek21}.

In this work, we take an approach towards temporal correlations using wiring operations. This allows the future measurement inputs to depend on the past measurement outputs. We use wirings to replace the problematic non-invasiveness measurability assumption by what we call \textit{retrievability of information}. Our proposed models form a strict generalisation of macrorealistic ones, which provides various advantages. First, as we allow disturbance, we argue that within the quantum formalism one only needs to consider the very basic Lüders instrument, which is present in every quantum measurement \cite{Pello4}. Second, as wiring allows for compensating measurements, the simplest cases of invasiveness, such as intentional rotation or translation, do not violate our models, as they can be compensated for on a later time-step. Third, we show that within quantum theory our models have a natural connection to joint measurability of quantum measurements \cite{heinosaari16b,JMreview}. This allows one to use the Busch-Lahti-Werner uncertainty relations \cite{buschrmp2014} as a justification for the fact that one is performing the optimal retrieving measurement. This further carries a basic property of macrorealistic models from Hermitian observables to the level of generalised measurements. Namely, macrorealistic models hold when a Hermitian observable is measured directly after itself. This is not the case for generalised measurements \cite{uola19a}. However, as any generalised measurement is jointly measurable with itself, our models do not allow a violation by simply asking the same question twice. Finally, we show that a quantum mechanically optimal protocol for testing our model is within the reach of today's experimental techniques, and implement the protocol on a photonic platform. We report that quantum theory is more general than the proposed classical retrieving models.

\textit{Hidden variable models for temporal correlations.---} The basic hidden variable model for temporal correlations is that of Leggett and Garg. In their seminal paper \cite{leggett85}, the idea of macrorealistic models was introduced. These models assume the existence of a macroscopic state $\lambda$ of the system, i.e.\ a variable that encodes the answer to possible measurements, and the possibility of measuring this state without disturbing it or its subsequent dynamics. The first one of these assumptions is called macrorealism per se (MRps for short) and the second one is called non-invasive measurability (NIM for short). The conjunction of these assumptions leads to what is called macroscopic hidden variable models, which again leads to the famous Leggett-Garg inequalities.

Let us first concentrate on a setting with two time steps, on each of which one either performs or does not perform a measurement. In this case, the MRps assumption implies that the statistics are given by
\begin{align}\label{Eq:MRps}
p(a,b|x,y)=\sum_\lambda p(\lambda)p(a,b|x,y,\lambda).
\end{align}
In other words, there exists a distribution $p(\lambda)$ of macrostates $\lambda$ and they give responses $a$ and $b$ to questions $x$ (on the first time step) and $y$ (on the second time step) according to the distribution $p(a,b|x,y,\lambda)$. It is important to note that such model exists for quantum theory by identifying the macrostates as pure quantum states. Introducing the NIM assumption, one gets more structure on the response functions, and a contradiction with quantum theory. The NIM assumption implies that $p(a,b|x,y,\lambda)=p(a|x,\lambda)p(b|y,\lambda)$. This is a strong requirement, and it was shown in Refs.~\cite{clemente16,clemente15,Kofler13} that in our scenario the conjunction of MRps and NIM is equivalent to the fact that the original distribution is non-signalling in both directions, i.e.
\begin{align}
    \sum_a p(a,b|x,y)&=p(b|y)\label{NSIT1}\\
    \sum_b p(a,b|x,y)&=p(a|x)\label{NSIT2}.
\end{align}
It is worth noting that the second one of these conditions is trivially fulfilled, as the choice of a future measurement $y$ can not affect the statistics of a past measurement $x$. The fact that quantum theory violates the conjunction of MRps and NIM can be seen through violations of Leggett-Garg inequalities \cite{Emary13}, or by the fact that quantum theory allows inherently disturbing measurements \cite{busch16,heinosaari10,uola19a} violating Eq.~(\ref{NSIT1}).

We are ready to present our proposal for a generalisation of the NIM assumption. We allow for the possibility of choosing the measurement performed on the second time step depending on the outcome of the first one. We call our assumption \textit{retrievability of information} (RoI for short) and state it on the level of macrostates mathematically as
\begin{align}\label{Eq:Retrieving}
    \sum_{\lambda} p(\lambda)p(b|0,y,\lambda)=\sum_{a,\lambda} p(\lambda)p(a,b|x,y_a,\lambda).
\end{align}
Here $(0,y)$ refers to the sequence with no measurement on the first time step and measurement $y$ on the final time step. Respectively, $(x,y_a)$ refers to measurement $x$ taking place on the first time step, whose output $a$ is then used to choose the input $y_a$ of the final measurement. In other words, the retrievability of information assumption states that the macrostate's response (on average) to the final measurement, when no measurement took place on the first step, can be retrieved (on average) by treating the output of the first measurement as an input on the final time step. This strictly includes scenarios covered by the NIM assumption. We summarise this section into a testable criterion as follows.
\begin{observation}
The conjunction of MRps and RoI leads to a model for the sequential statistics that fulfills
\begin{align}\label{Eq:RoI}
    \sum_a p(a,b|0,y)=\sum_a p(a,b|x,y_a).
\end{align}
\end{observation}
The proof of this statement is a straight-forward implication of Eq.~(\ref{Eq:Retrieving}). We present the details in the Appendix.

\textit{Retrieving in the quantum scenario.---} In quantum theory, generalised measurements are given by positive-operator valued measures (POVM for short) acting on a finite-dimensional Hilbert space, i.e.\ a collection of positive-semidefinite matrices $(A_{a})_a$, where $a$ labels the outcome, for which $\sum_a A_{a}=\openone$. Such a POVM gives the measurement outcome probabilities according to the Born rule, i.e. in a given quantum state $\varrho$, that is a positive semi-definite unit-trace operator, one has the probability distribution $p(a|\varrho)=\text{tr}[A_{a}\varrho]$. The state update caused by such a measurements is given by a quantum instrument $(\mathcal I_{a})_a$, i.e.\ a collection of completely positive maps with the property that $\sum_a \mathcal I_{a}$ is a trace-preserving map (i.e.\ a quantum channel) and $\text{tr}[\mathcal I_{a}(\varrho)]=\text{tr}[A_{a}\varrho]$. Here $\mathcal I_{a}(\varrho)$ represents the sub-normalised post-measurement state when the measurement of $(A_{a})_a$ in the initial state $\varrho$ is performed and outcome $a$ is obtained. It is a central result of quantum measurement theory \cite{busch16,Pello4}, that every instrument that describes a given POVM $(A_{a})_a$ is of the form 
\begin{align}\label{Eq:Generalinstrument}
\mathcal I_{a}(\varrho)=\mathcal{E}_a(\sqrt{A_{a}}\varrho\sqrt{A_{a}}),
\end{align}
i.e.\ one first performs a minimally disturbing L\"uders state update and then applies a noise channel $\mathcal{E}_a$ depending on the obtained outcome $a$.

A natural division between classical and quantum properties of measurements is given by \textit{joint measurability} \cite{heinosaari16b,JMreview}. For a pair of POVMs $(A_{a})_a$ and $(B_{b})_b$, we ask whether there exists a third POVM $(G_{a,b})_{a,b}$ such that $A_{a}=\sum_b G_{a,b}$ for each $a$ and $B_{b}=\sum_a G_{a,b}$ for each $b$. If such POVM exists, the pair is called jointly measurable.

For our purposes, a central property of jointly measurable pairs of POVMs is that they are exactly the ones that allow a realisation in a sequence with the second measurement being a retrieving one \cite{heinosaari15b,Pello7,JMreview}. More precisely, the statistics of any $(B_{b})_b$ jointly measurable with a POVM $(A_{a})_a$ can be obtained from properly chosen post-measurement states $\mathcal I_{a}(\varrho)$. Instead of measuring $(B_{b})_b$, this typically requires one to make a different, so-called retrieving measurement $(\tilde B_{b})_b$. In general, such retrieving measurement can act on a larger Hilbert space \cite{HaHeMi2018}, and the joint measurement takes the form, cf. Eq.~(\ref{Eq:Generalinstrument})
\begin{align}
\text{tr}[G_{a,b}\varrho]&=\text{tr}[\tilde B_b\mathcal{E}_a(\sqrt{A_a}\varrho\sqrt{A_a})]\\
&=\text{tr}[\mathcal{E}_a^*(\tilde B_b)\sqrt{A_a}\varrho\sqrt{A_a}]\label{Eq:retrievinginstru}.
\end{align}
Here the star represents the Heisenberg picture and the noise channels $\mathcal{E}_a$ can possibly map between different Hilbert spaces.

The above Eq.~(\ref{Eq:retrievinginstru}) is important for us for three reasons. First, we note that the noise channels $\mathcal{E}_a$ can possibly map between different Hilbert spaces. However, as the Heisenberg picture maps back to the original space, we can define the POVMs $(\tilde B_{b|a})_b$ with $\tilde B_{b|a}:=\mathcal{E}_a^*(\tilde B_b)$. These act on the same space as $\varrho$. Hence, we do not require any additional degrees of freedom to implement such scenario. Second, the data of any POVM that is jointly measurable with $(A_a)_a$ is still in the system once one has performed the basic L\"uders state update, cf. \cite{heinosaari15b}. To get this data, one is simply required to perform a retrieving measurement, whose input can depend on the output of the first measurement, cf. $(\tilde B_{b|a})_b$ above. Finally, as the process for finding the retrieving measurement is constructive \cite{heinosaari15b,Pello7}, we argue that retrieving scenarios without the noise channels $\mathcal{E}_a$ are sufficient for the experimental setting considered here (noisy $Z$ and $X$ measurements), cf. Remark 2 of the Appendix. In other words, it is sufficient to consider only the L\"uders state update. In such case, the retrieving measurements do not need to depend on the output of the first measurement, i.e. $\tilde B_{b|a}=\mathcal{E}_a^*(\tilde B_b)=\tilde B_b$.

\begin{figure}[htbp]
\begin{center}
\includegraphics[width=0.48\textwidth]{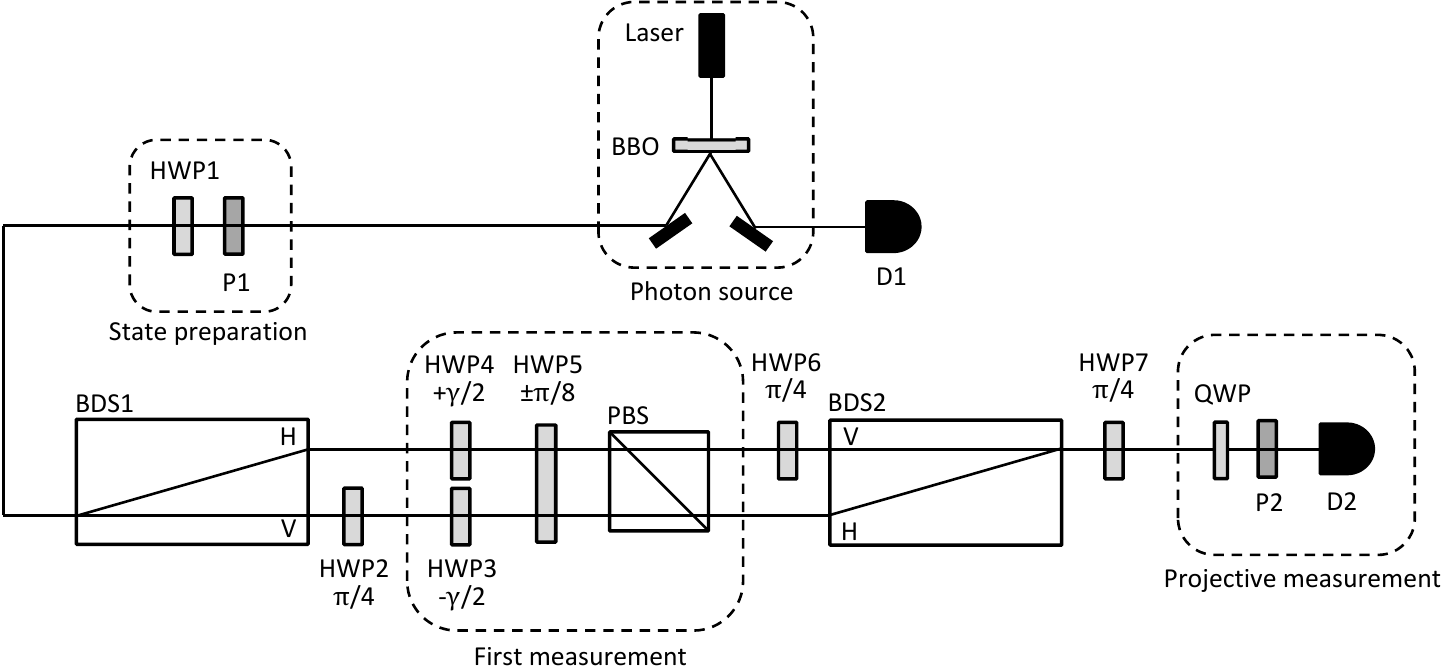} 
\caption{Experimental setup.}\label{Fig:ExpSetup}
\end{center}
\end{figure}

\textit{Experimental setup.---} The photon source is a type-I beta-barium borate crystal pumped with a tightly focused continuous wave laser of the wavelength 405 nm. The crystal produces a pair of photons (the signal and idler) around the wavelength of 810 nm through spontaneous down-conversion process. The idler photon is registered by the single photon detector D1, which triggers the coincidence electronics and data collection. The initial state for measurements is prepared by passing the signal photon through a fixed polarizer P1 and rotated half-wave plate HWP1. Hence, we prepare the superpositions of the polarization states $|H\rangle$ and $|V\rangle$.
In the beam displacer BDS1 the horizontal and vertical polarization components are separated and after HWP2 the polarization information is encoded into degree of freedom of the optical path, the upper and lower arm, cf. Ref.~\cite{Chen2019}. The measurement of the first POVM $(A^\gamma_+,A^\gamma_-)$
is performed with the help of the half-wave plates HWP3 (the rotation angle $-\gamma/2$), HWP4 ($+\gamma/2$), HWP5 ($+\pi/8$) and the polarizing beam splitter PBS.
The transmitted (resp.\ reflected) beams of PBS correspond to the outcome $+1$ (resp.\ $-1$).
By changing the parameter $\gamma$ we obtain different POVMs. 
In order to simplify the setup, only the transmission arm of the PBS is used, and the reflective version of measurement is simulated by HWP5 with the rotation angle of $-\pi/8$. The information of the system that is encoded in the path basis is recoded back to the polarization basis by recombining the optical paths via HWP6, BDS2 and HWP7. The final projective measurement is performed with the quarter-wave plate QWP, the polarizer P2 and the detector D2. The experimental setup is depicted in Fig.~\ref{Fig:ExpSetup}.

For $0\le\gamma\le\pi/4$,
the first POVM $A^\gamma_\pm
=\frac12\big[\id\pm\sin{(2\gamma)}\sigma_z\big]$ is a noisy version of the sharp $Z$ measurement
and the implemented state update is given by the Lüders instrument $\mathcal I_\pm^\gamma(\varrho)=\sqrt{A^\gamma_\pm}\varrho\sqrt{A^\gamma_\pm}$ where $\varrho=|\Psi_{\rm in}\rangle\langle\Psi_{\rm in}|$ is the initial state, cf. the Appendix for a detailed derivation.
Note that $A^0_\pm
=\frac12\id$ and 
$\mathcal I_\pm^0(\varrho)=\frac12\varrho$ so the case $\gamma=0$ can be interpreted as if there is no first measurement at all. The maximally disturbing first measurement (i.e.\ state diagonalization) is obtained when $\gamma=\pi/4$ and $A_\pm^{\pi/4}=Z_\pm$. We present below three different scenarios for probing temporal correlations with our setup.

\begin{figure}[htbp]
\begin{center}
\includegraphics[width=0.45\textwidth]{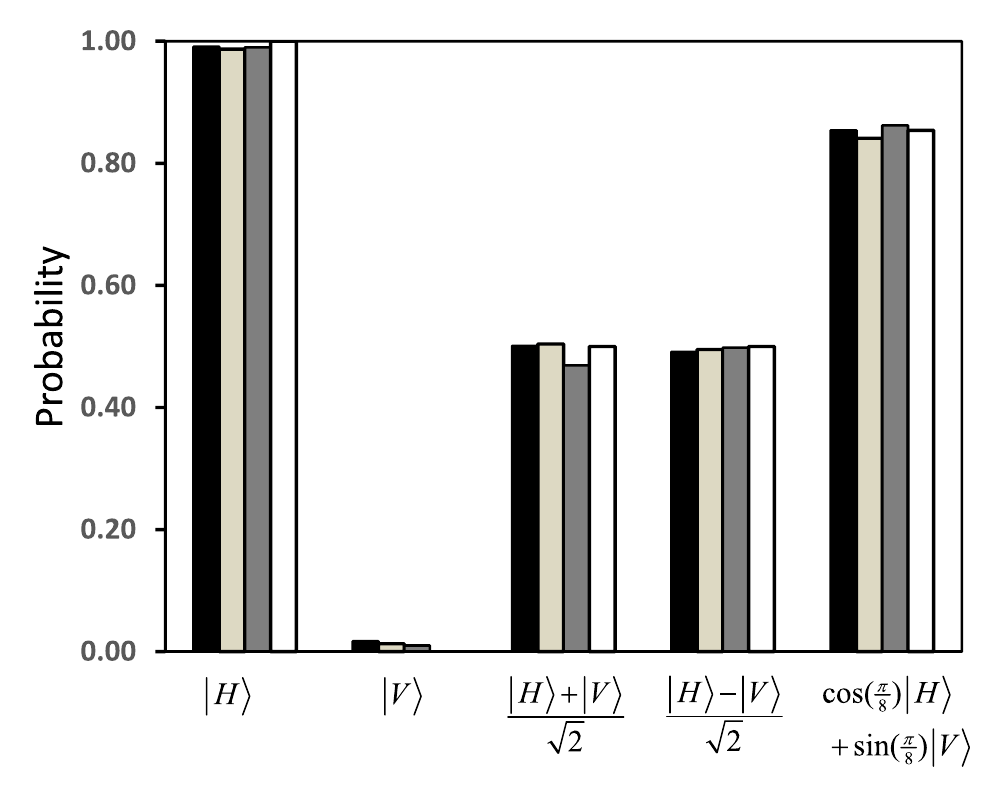} 
\caption{Macrorealistic situation. The final measurement is  sharp $Z$. On the $x$-axis, we have the input state $|\Psi_{\rm in}\rangle$. In the Figure we have plotted the experimental probabilities of the plus outcome of the second margin for $\gamma=0$ (black), $\gamma=\pi/8$ (light gray), and $\gamma=\pi/4$ (gray). The white bar represents the theoretical values which equal $\langle\Psi_{\rm in}|Z_+\Psi_{\rm in}\rangle=\big|\langle H|\Psi_{\rm in}\rangle\big|^2$ for all values of $\gamma$.}
\label{kuva2}
\end{center}
\end{figure}

\textit{The macrorealistic case.---} In the simplest case, we have a sharp (or noisy) $Z$ as the first measurement and a sharp $Z$ as the final measurement. In this case, the sequence generates a joint POVM $G^\gamma_{a,b}=\sqrt{A^\gamma_a}Z_b\sqrt{A^\gamma_a}$. The second margin of this POVM is $Z_\pm$ which does not depend on $\gamma$. This corresponds to a macrorealistic scenario, i.e. the Eq.~(\ref{NSIT1}) holds with $x=A^\gamma$ and $y=Z$. This is demonstrated in Fig.~\ref{kuva2}

\begin{figure}[htbp]
\begin{center}
\includegraphics[width=0.45\textwidth]{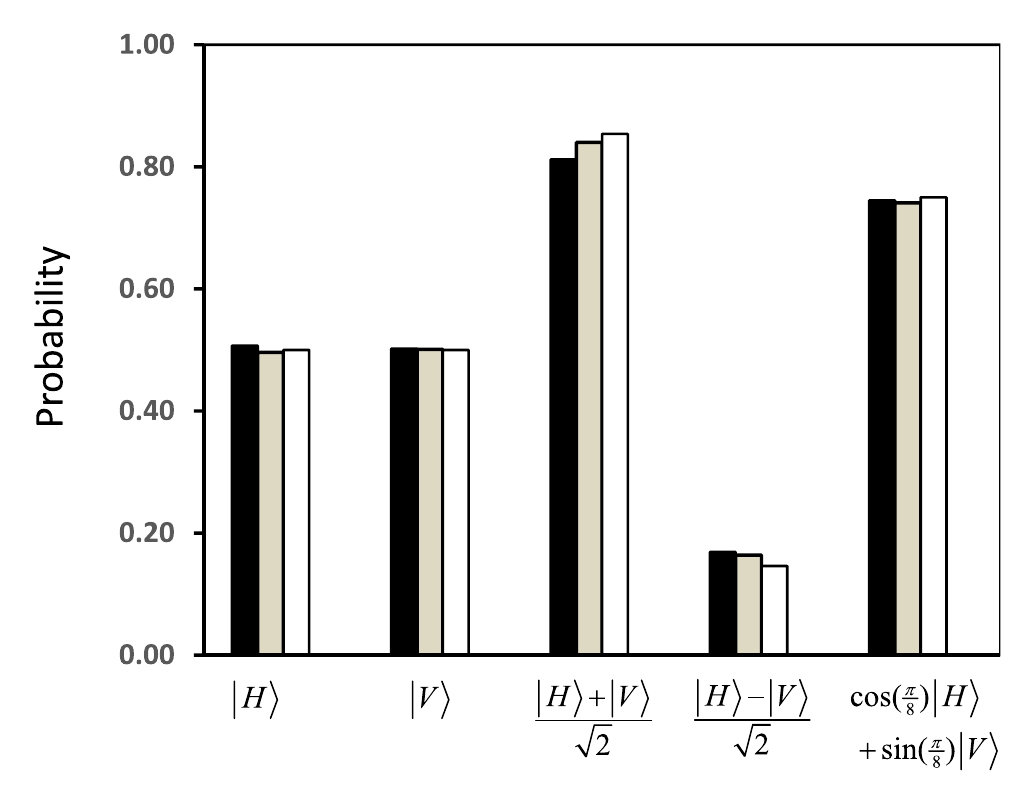} 
\caption{Retrieving situation. The black bar represents the experimental data of the plus outcome of noisy $X$ 
with no first measurement. The light gray bar is the data of sharp $X$ after the measurement of noisy $Z$. The white bar is the theoretical prediction $\langle\Psi_{\rm in}|B^{\pi/8}_+\Psi_{\rm in}\rangle$, which is the same for both cases.}
\label{kuva3}
\end{center}
\end{figure}

\textit{The retrieving case.---} If the final projective (sharp) measurement is $\tilde B_\pm=X_\pm=\frac12(\id\pm\sigma_x)$ then the sequence generates a joint POVM
\begin{eqnarray}
G^\gamma_{a,b}&=&\mathcal I^{\gamma\,*}_a(X_b)=
\sqrt{A_a^\gamma}X_b\sqrt{A_a^\gamma} \\
&=&
\frac{1}{4}\big[\id+a\sin{(2\gamma)}\sigma_z+b\cos{(2\gamma)}\sigma_x\big].
\end{eqnarray}
The first margin is naturally the POVM $A_\pm^\gamma$ measured first and the second margin is a noisy $X$-observable given by $B^\gamma_\pm=\frac12\big[\id\pm\cos{(2\gamma)}\sigma_x\big]$. According to \cite{BLW2014}, the minimum value $2(2-\sqrt{2})$ for the Busch-Lahti-Werner uncertainty sum $\Delta(\ms A,\ms Z)^2+\Delta(\ms B,\ms X)^2$ is reached for $\ms A=\ms A^{\pi/8}$ and $\ms B=\ms B^{\pi/8}$ where $\Delta$ is the worst-case Wasserstein-2 distance for POVMs, cf.\ Appendix. In this sense, our measurement setting with $\gamma=\pi/8$ realizes the optimal approximate joint measurement of $\ms Z$ and $\ms X$ in a sequence; in this case
\begin{align}
A^{\pi/8}_\pm
=\frac12\Big(\id\pm\frac{1}{\sqrt2}\sigma_z\Big),\quad
B^{\pi/8}_\pm=\frac12\Big(\id\pm\frac{1}{\sqrt2}\sigma_x\Big).
\end{align}
When $\gamma=\frac{\pi}{8}$ then the minimum $\frac32$ of the uncertainty sum ${\rm Var}(A^{\pi/8},\varrho)+{\rm Var}(B^{\pi/8},\varrho)$ is reached for all pure states $\varrho=|\Psi_{\rm in}\rangle\langle\Psi_{\rm in}|$ where
$|\Psi_{\rm in}\rangle$ is any {\it real} superposition of the basis vectors.
Hence, it is reasonable to prepare only these kinds of pure states.
We further note that the correlation ${\rm Corr}(G^{\pi/8},\varrho)$ between the margins of the joint measurement $G^{\pi/8}$ in the state  $\varrho$ satisfies $-\frac13\le{\rm Corr}(G^{\pi/8},\varrho)\le\frac13$,
where the bounds $\pm\frac13$ are reached with pure states $\varrho=|\Psi_{\rm in}^\pm\rangle\langle\Psi_{\rm in}^\pm|$ where
\begin{eqnarray}
|\Psi_{\rm in}^\pm\rangle=\cos(\pi/8)\ket H\mp\sin(\pi/8)\ket V.
\end{eqnarray}
see Appendix for detailed calculations. We take the maximally anti-correlated state $|\Psi_{\rm in}^-\rangle$ as one of our test states in addition to the eigenstates of $Z$ and $X$. We have presented the realised retrieving scenario in Fig.~\ref{kuva3} for the case where there is no first measurement and the final one is $B^{\pi/8}$, and the first one being $A^{\pi/8}$ and the final $X$. We note that this is in line with Eq.~(\ref{Eq:RoI}), but the corresponding macrorealistic scenario, where one is forced to measure noisy $X$ as the final measurement in both sequences, breaks the condition on macrorealism given by Eq.~(\ref{NSIT1}).

\textit{The no-retrieving case.---} We have seen above that the least noisy version of $X$ that can be retrieved after the measurement of $A^{\pi/8}$ with the above optimal retrieving scenario, is $B^{\pi/8}$. Hence, if we simply choose a setup, where we require the retrieving of sharp $X$, the data violates the retrieving condition of Eq.~(\ref{Eq:RoI}), cf. Fig.~\ref{kuva4}. Indeed, if one wants to retrieve sharp $X$, this forces the first measurement to be trivial, see Appendix.

\begin{figure}[htbp]
\begin{center}
\includegraphics[width=0.45\textwidth]{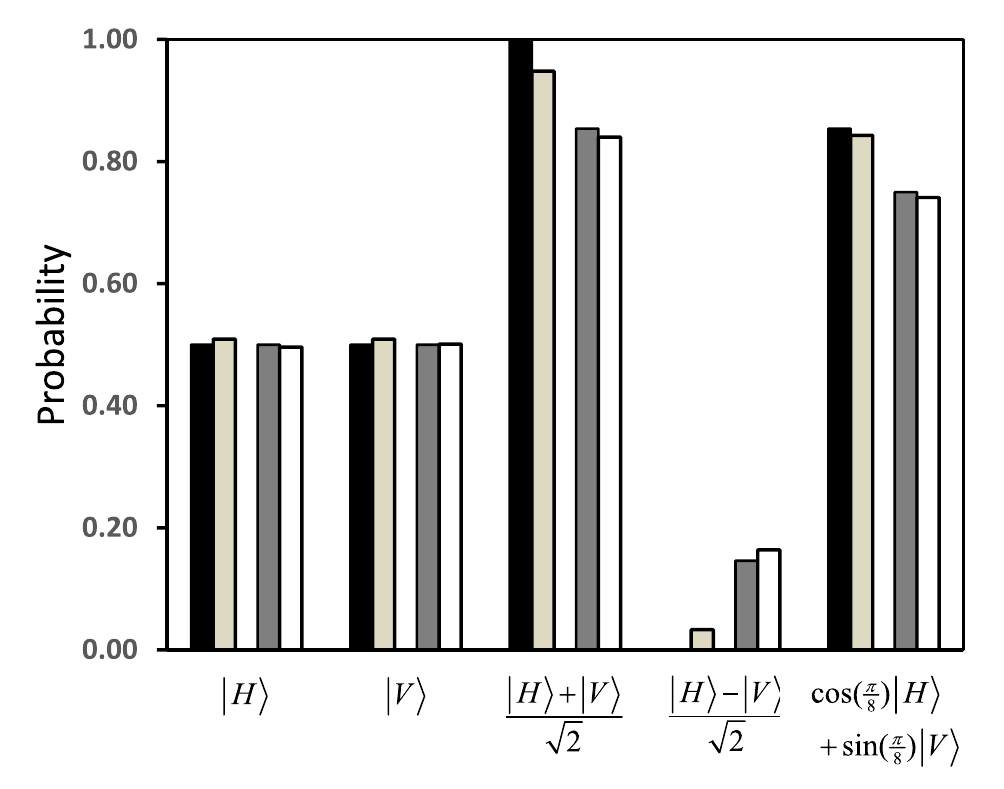} \caption{No-retrieving situation.
The black bar represents the theoretical prediction for sharp $X$ when there is no first measurement, i.e. $\langle\Psi_{\rm in}|X_+\Psi_{\rm in}\rangle$, and the light gray is its experimental value. The dark gray bar represents the theoretical values for $\langle\Psi_{\rm in}|B^{\pi/8}_+\Psi_{\rm in}\rangle$, and the white bar is its experimental value. As the bars differ for the last three states, and as the retrieving protocol is optimal according to quantum theory, one can conclude that the retrieving of sharp $X$ is not possible.
}
\label{kuva4}
\end{center}
\end{figure}

\textit{Conclusions.---} We have proposed a generalisation of macrorealistic hidden variable models. Our models replace the non-invasive measurability postulate by a more general retrievability of information assumption. Within the quantum formalism, we have demonstrated that this eases the clumsiness loophole associated with macrorealism in the sense, that we were able to narrow down the required measurement procedures in general into only the L\"uders instrument, which is present in every quantum measurement. To complete the retrieving protocol, we have focused on the qubit case, where we have connected the problem of finding the best possible retrieving measurement into the fundamental precision limit of quantum theory given by the Busch-Lahti-Werner uncertainty relation. We have experimentally demonstrated an optimal retrieving protocol, and reported a violation of the associated hidden variable models. Hence, any hidden variable model allowing for retrieving in our scenario would need to go beyond the fundamental quantum precision limit.

For future research, it will be of interest to investigate our models in the framework of more measurements and different time evolutions. We expect the former to find interesting connections with recent advances on error-disturbance uncertainty relations \cite{Yu22a,Yu22b}. For the latter, we note that closed system time evolutions are implicitly present in our protocol by using the Heisenberg picture. However, the use of open system dynamics will pose an interesting constraint, as some information can leak to the environment and, hence, not be retrievable. In this case, not all jointly measurable pairs can be retrieved. We believe that this aspect of incompatible measurements in open quantum systems is of interest on its own and should be further investigated. Another interesting direction is to look into scenarios where the retrieving step requires an adaptive feed-forward mechanism, i.e. when the second measurement's input depends on the output of the first one. We believe that these can find applications in the context of Busch-Lahti-Werner uncertainty relations, as well as in recently found applications of sequential joint measurements in quantum thermodynamics \cite{Beyer22}. Finally, a possible line of research is to investigate further relaxations of the retrievability assumption. This could have intriguing implications on the clumsiness loophole.

\textit{Acknowledgements.---} We would like to thank Costantino Budroni for useful discussions on temporal correlations. R.U.\ is thankful for the financial support from the Swiss National Science Foundation (Ambizione PZ00P2-202179). E.H.\ is funded by the National Research Foundation, Prime Minister's Office, Singapore and the Ministry of Education, Singapore under the Research Centres of Excellence programme.

\bibliographystyle{unsrt}
\bibliography{references}

\begin{thebibliography}{10}

\bibitem{leggett85}
A.~J. Leggett and Anupam Garg.
\newblock Quantum mechanics versus macroscopic realism: Is the flux there when
  nobody looks?
\newblock {\em Phys. Rev. Lett.}, 54:857--860, Mar 1985.

\bibitem{Emary13}
Clive Emary, Neill Lambert, and Franco Nori.
\newblock Leggett–garg inequalities.
\newblock {\em Reports on Progress in Physics}, 77(1):016001, Dec 2013.

\bibitem{wilde11}
Mark~M. Wilde and Ari Mizel.
\newblock Addressing the clumsiness loophole in a leggett-garg test of
  macrorealism.
\newblock {\em Found. Phys.}, 42:256--265, Sep 2011.

\bibitem{uola19a}
R.~{Uola}, G.~{Vitagliano}, and C.~{Budroni}.
\newblock Leggett-garg macrorealism and the quantum nondisturbance conditions.
\newblock {\em Phys. Rev. A}, 100:042117, 2019.

\bibitem{george13}
Richard~E. George, Lucio~M. Robledo, Owen J.~E. Maroney, Machiel~S. Blok,
  Hannes Bernien, Matthew~L. Markham, Daniel~J. Twitchen, John J.~L. Morton,
  G.~Andrew~D. Briggs, and Ronald Hanson.
\newblock Opening up three quantum boxes causes classically undetectable
  wavefunction collapse.
\newblock {\em Proc. Natl. Acad. Sci.}, 2013.

\bibitem{knee16}
George~C. Knee, Kosuke Kakuyanagi, Mao-Chuang Yeh, Yuichiro Matsuzaki, Hiraku
  Toida, Hiroshi Yamaguchi, Shiro Saito, Anthony~J. Leggett, and William~J.
  Munro.
\newblock A strict experimental test of macroscopic realism in a
  superconducting flux qubit.
\newblock {\em Nat. Commun.}, 7:13253, Nov 2016.

\bibitem{Schmid22}
David Schmid.
\newblock Macrorealism as strict classicality in the framework of generalized
  probabilistic theories (and how to falsify it), 2022.

\bibitem{Grabowecky22}
Michael~J. Grabowecky, Christopher A.~J. Pollack, Andrew~R. Cameron, Robert~W.
  Spekkens, and Kevin~J. Resch.
\newblock Experimentally bounding deviations from quantum theory for a photonic
  three-level system using theory-agnostic tomography.
\newblock {\em Physical Review A}, 105(3), mar 2022.

\bibitem{Mazurek21}
Michael~D. Mazurek, Matthew~F. Pusey, Kevin~J. Resch, and Robert~W. Spekkens.
\newblock Experimentally bounding deviations from quantum theory in the
  landscape of generalized probabilistic theories.
\newblock {\em PRX Quantum}, 2:020302, Apr 2021.

\bibitem{Pello4}
Juha-Pekka Pellonp\"a\"a.
\newblock Quantum instruments: {II}. measurement theory.
\newblock {\em J. Phys. A}, 46(2):025303, dec 2013.

\bibitem{heinosaari16b}
T.~Heinosaari, T.~Miyadera, and M.~Ziman.
\newblock An invitation to quantum incompatibility.
\newblock {\em J. Phys. A}, 49:123001, 2016.

\bibitem{JMreview}
Otfried G\"uhne, Erkka Haapasalo, Tristan Kraft, Juha-Pekka Pellonp\"a\"a, and
  Roope Uola.
\newblock Incompatible measurements in quantum information science.
  arxiv:2112.06784.

\bibitem{buschrmp2014}
Paul Busch, Pekka Lahti, and Reinhard~F. Werner.
\newblock Colloquium: Quantum root-mean-square error and measurement
  uncertainty relations.
\newblock {\em Rev. Mod. Phys.}, 86:1261--1281, 2014.

\bibitem{clemente16}
Lucas Clemente and Johannes Kofler.
\newblock No fine theorem for macrorealism: Limitations of the leggett-garg
  inequality.
\newblock {\em Phys. Rev. Lett.}, 116(15):150401, Apr 2016.

\bibitem{clemente15}
Lucas Clemente and Johannes Kofler.
\newblock Necessary and sufficient conditions for macroscopic realism from
  quantum mechanics.
\newblock {\em Phys. Rev. A}, 91:062103, Jun 2015.

\bibitem{Kofler13}
Johannes Kofler and \ifmmode \check{C}\else~\v{C}\fi{}aslav Brukner.
\newblock Condition for macroscopic realism beyond the leggett-garg
  inequalities.
\newblock {\em Phys. Rev. A}, 87:052115, May 2013.

\bibitem{busch16}
P.~{Busch}, P.~{Lahti}, J.-P. {Pellonp\"a\"a}, and K.{Ylinen}.
\newblock {\em Quantum Measurement (Theoretical and Mathematical Physics)}.
\newblock Springer, 2016.

\bibitem{heinosaari10}
T.~Heinosaari and M.M. Wolf.
\newblock Non-disturbing quantum measurements.
\newblock {\em J. Math. Phys.}, 51:092201, 2010.

\bibitem{heinosaari15b}
Teiko Heinosaari and Takayuki Miyadera.
\newblock Universality of sequential quantum measurements.
\newblock {\em Phys. Rev. A}, 91:022110., Feb 2015.

\bibitem{Pello7}
Erkka Haapasalo and Juha-Pekka Pellonp\"a\"a.
\newblock Optimal quantum observables.
\newblock {\em J. Math. Phys.}, 58(12):122104, 2017.

\bibitem{HaHeMi2018}
E.~Haapasalo, T.~Heinosaari, and T.~Miyadera.
\newblock The unavoidable information flow to environment in quantum
  measurements.
\newblock {\em J. Math. Phys.}, 59:082106, Aug 2018.

\bibitem{Chen2019}
Jiang-Shan Chen, Meng-Jun Hu, Xiao-Min Hu, Bi-Heng Liu, Yun-Feng Huang,
  Chuan-Feng Li, Can-Guang Guo, and Yong-Sheng Zhang.
\newblock Experimental realization of sequential weak measurements of
  non-commuting pauli observables.
\newblock {\em Optics Express}, 27(5):6089, feb 2019.

\bibitem{BLW2014}
Paul Busch, Pekka Lahti, and Reinhard~F. Werner.
\newblock Heisenberg uncertainty for qubit measurements.
\newblock {\em Phys. Rev. A}, 89:012129, 2014.

\bibitem{Yu22a}
Sixia Yu, Ya-Li Mao, Chang Niu, Hu~Chen, Zheng-Da Li, and Jingyun Fan.
\newblock Measurement uncertainty relation for three observables, 2022.

\bibitem{Yu22b}
Ya-Li Mao, Hu~Chen, Chang Niu, Zheng-Da Li, Sixia Yu, and Jingyun Fan.
\newblock Testing heisenberg's measurement uncertainty relation of three
  observables, 2022.

\bibitem{Beyer22}
Konstantin Beyer, Roope Uola, Kimmo Luoma, and Walter~T. Strunz.
\newblock Joint measurability in nonequilibrium quantum thermodynamics.
\newblock {\em Physical Review E}, 106(2), 2022.

\end{thebibliography}

\appendix

\onecolumngrid
\vspace{0.5cm}
\section*{Appendix}
\vspace{0.5cm}

\section{Theoretical details on classical models}

In the main text, we claim that the conjunction of MRps and RoI leads to sequential statistics that fulfill
\begin{align}\label{Eq:RoI2}
    \sum_a p(a,b|0,y)=\sum_a p(a,b|x,y_a).
\end{align}

To prove this, we note that MRps together with the Bayes rule and RoI implies
\begin{align*}
    \sum_a p(a,b|0,y)&=\sum_{a,\lambda} p(\lambda)p(a,b|0,y,\lambda)\\
    &=\sum_{a,\lambda} p(\lambda)p(a|b,0,y,\lambda)p(b|0,y,\lambda)\\
    &=\sum_\lambda p(\lambda)p(b|0,y,\lambda)\\
    &=\sum_{a,\lambda} p(\lambda)p(a,b|x,y_a,\lambda)\\
    &=\sum_{a} p(a,b|x,y_a).
\end{align*}
We note that this condition forms one type of an analogue of Eq.~(\ref{NSIT1}) to our general setting. Namely, whereas Eq.~(\ref{NSIT1}) and Eq.~(\ref{NSIT2}) fully characterise the conjunction of MRps and NIM, Eq.~(\ref{Eq:RoI2}) fully characterises the correlations given by the RoI assumption when MRps is assumed. The last claim is implied by the above chain of equations.

\subsection{Quantum model for classical retrievable correlations}

Assume that a correlation table allows a hidden variable model satisfying macrorealism per se and RoI. Take a Hilbert space with an orthonormal basis $\{\ket\lambda\}_\lambda$
and define a state $\varrho:=\sum_\lambda p(\lambda)\kb\lambda\lambda$ and a (joint) POVM $  G_{a,b}:=\sum_\lambda p(a,b|x,y_a,\lambda)\kb\lambda\lambda$. This gives $\sum_a\tr[\varrho   G_{a,b}]=\sum_{a,\lambda}p(\lambda)p(a,b|x,y_a,\lambda)=
\sum_{\lambda}p(\lambda)p(b|0,y,\lambda)=\tr[\varrho   B_{b}]$
where the effects $B_{b}:=\sum_\lambda p(b|0,y,\lambda)\kb\lambda\lambda$ constitute a POVM. Since all operators above are diagonal and $\varrho$ is unknown,
the equation $\sum_a\tr[\varrho   G_{a,b}]=\tr[\varrho   B_{b}]$ holds for all states $\varrho$ implying $\sum_a   G_{a,b}=  B_b$. 
Hence, any hidden variable model satisfying macrorealism per se and RoI has a quantum realisation using a diagonal state and a joint measurement. As any joint measurement can be realised in a sequence with retrieving measurements \cite{heinosaari15b,Pello7}, this gives a sequential quantum model.

\subsection{Classical model for quantum retrievable measurements}

Suppose that
$$
\text{tr}[G_{a,b}\varrho]=\text{tr}[\tilde B_b\mathcal{E}_a(\sqrt{A_a}\varrho\sqrt{A_a})]=\text{tr}[\mathcal{E}_a^*(\tilde B_b)\sqrt{A_a}\varrho\sqrt{A_a}].
$$
Let the variables $\lambda$ be the unit vectors of the Hilbert space. By using the spectral decomposition, one can write $\varrho=\sum_\lambda p(\lambda)\kb\lambda\lambda$ (where only countably many probabilities $p(\lambda)$ are nonzero)
so 
\begin{eqnarray*}
\sum_{\lambda}p(\lambda)p(b|0,y,\lambda)
&=&\sum_{\lambda}p(\lambda)\langle\lambda|B_b|\lambda\rangle
=\text{tr}[B_{b}\varrho]=\sum_a \text{tr}[G_{a,b}\varrho]
=\sum_a\text{tr}[\mathcal{E}_a^*(\tilde B_b)\sqrt{A_a}\varrho\sqrt{A_a}] \\
&=&\sum_{a,\lambda}p(\lambda)\langle\lambda|\sqrt{A_a}\mathcal{E}_a^*(\tilde B_b)\sqrt{A_a}|\lambda\rangle  =\sum_{a,\lambda}p(\lambda)p(a,b|x,y_a,\lambda).
\end{eqnarray*}
Here $x=A$, $y_a=\mathcal{E}_a^*(\tilde B)$, and $y=B$.

\section{Experimental details}

We present some details of the experimental setup and both theoretical and experimental results.

\begin{figure}[h!]
\centering
\includegraphics[scale=1]{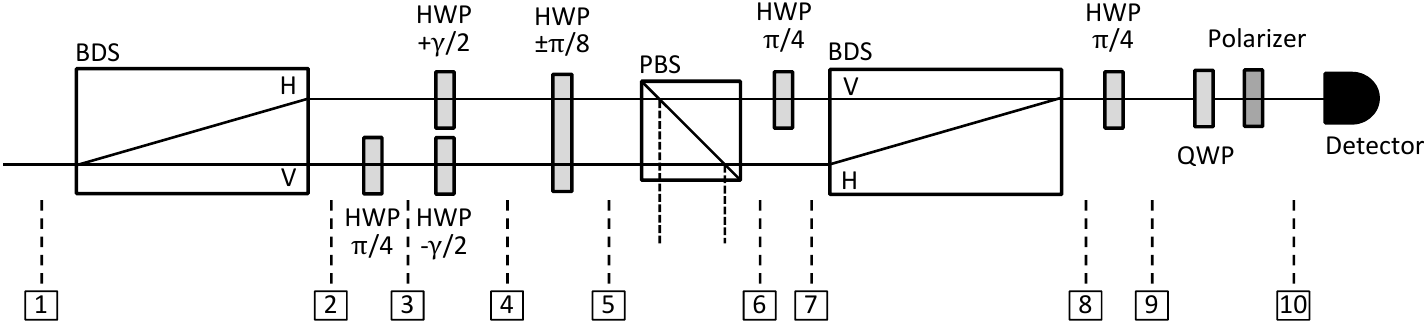}
\caption{Experimental setup}
\end{figure}

Let $\{\ket H,\ket V\}$ be the polarization basis and define the Pauli matrices
$\sigma_x:=|H\>\<V|+|V\>\<H|$,
$\sigma_y:=-i|H\>\<V|+i|V\>\<H|$, and
$\sigma_z:=|H\>\<H|-|V\>\<V|$. We consider all operators as matrices in the basis $\{\ket H,\ket V\}$.

\noindent $\boxed{1}$ The initial (input) pure state:
\begin{equation*}
\Psi_{\rm in}=\alpha \vert H\rangle + \beta \vert V\rangle 
\end{equation*}
where the complex numbers $\alpha$ and $\beta$ satisfy $\vert\alpha\vert^2+\vert\beta\vert^2=1$.\\

\noindent $\boxed{2}$ The beam displacer separates the horizontal and vertical polarization components:
\begin{equation*}
\Psi_2=\alpha \vert 0\rangle \otimes \vert H\rangle + \beta \vert 1\rangle\otimes \vert V\rangle
\end{equation*}
The path states $\vert 0\rangle$ and $\vert 1\rangle$ represent photons flying along the up arm and down arm.\\

\noindent $\boxed{3}$ Vertical state flip:
\begin{equation*}
\Psi_3=\alpha \vert 0\rangle \otimes \vert H\rangle + \beta \vert 1\rangle\otimes \vert H\rangle
\end{equation*}

\noindent $\boxed{4}$ Phase shift on both parts:
\begin{equation*}
\Psi_4=\alpha \vert 0\rangle\otimes \left[ \cos(\gamma)\vert H\rangle+\sin(\gamma)\vert V\rangle \right]+\beta \vert 1\rangle\otimes \left[ \cos(\gamma)\vert H\rangle-\sin(\gamma)\vert V\rangle \right]
\end{equation*}

\noindent $\boxed{5}$ Rotation with $\text{HWP}(\phi)$ (we will use rotation angles $\phi=\pm\pi/8$):
\begin{equation*}
\begin{aligned}
\Psi_5=\alpha \vert 0\rangle &\otimes \left\lbrace \cos(\gamma)\left[\cos(2\phi)\vert H\rangle + \sin(2\phi)\vert V\rangle \right] + \sin(\gamma)\left[\sin(2\phi)\vert H\rangle - \cos(2\phi)\vert V\rangle \right]  \right\rbrace\\ 
+ \beta \vert 1\rangle &\otimes \left\lbrace \cos(\gamma)\left[\cos(2\phi)\vert H\rangle + \sin(2\phi)\vert V\rangle \right] - \sin(\gamma)\left[\sin(2\phi)\vert H\rangle - \cos(2\phi)\vert V\rangle \right]  \right\rbrace
\end{aligned}
\end{equation*}

\noindent $\boxed{6}$ Only the horizontal component is transmitted through the polarizing beam splitter:
\begin{equation*}
\begin{aligned}
\Psi_6=\alpha \vert 0\rangle &\otimes\left[ \cos(\gamma)\cos(2\phi)+\sin(\gamma)\sin(2\phi)\right]\vert H\rangle\\
+\beta \vert 1\rangle &\otimes\left[ \cos(\gamma)\cos(2\phi)-\sin(\gamma)\sin(2\phi)\right]\vert H\rangle
\end {aligned}
\end{equation*}

\noindent $\boxed{7}$ Horizontal state flip:
\begin{equation*}
\begin{aligned}
\Psi_7=\alpha \vert 0\rangle &\otimes\left[ \cos(\gamma)\cos(2\phi)+\sin(\gamma)\sin(2\phi)\right]\vert V\rangle\\
+\beta \vert 1\rangle &\otimes\left[ \cos(\gamma)\cos(2\phi)-\sin(\gamma)\sin(2\phi)\right]\vert H\rangle
\end {aligned}
\end{equation*}

\noindent $\boxed{8}$ The polarization components are merged in the beam displacer:
\begin{equation*}
\begin{aligned}
\Psi_8=\alpha&\left[ \cos(\gamma)\cos(2\phi)+\sin(\gamma)\sin(2\phi)\right]\vert V\rangle\\
+\beta &\left[ \cos(\gamma)\cos(2\phi)-\sin(\gamma)\sin(2\phi)\right]\vert H\rangle
\end {aligned}
\end{equation*}

\noindent $\boxed{9}$ The polarization components are swapped:
\begin{equation*}
\begin{aligned}
\Psi_9=\alpha&\left[ \cos(\gamma)\cos(2\phi)+\sin(\gamma)\sin(2\phi)\right]\vert H\rangle\\
+\beta &\left[ \cos(\gamma)\cos(2\phi)-\sin(\gamma)\sin(2\phi)\right]\vert V\rangle
\end {aligned}
\end{equation*}

\noindent Especially, for the angles $\phi=\pm\pi/8$:
\begin{equation*}
\begin{aligned}
\Psi_9^+&=\dfrac{\sqrt{2}}{2}\alpha\left[\cos(\gamma)+\sin(\gamma)\right]\vert H\rangle +\dfrac{\sqrt{2}}{2}\beta\left[\cos(\gamma)-\sin(\gamma)\right]\vert V\rangle  \\
\Psi_9^-&=\dfrac{\sqrt{2}}{2}\alpha\left[\cos(\gamma)-\sin(\gamma)\right]\vert H\rangle +\dfrac{\sqrt{2}}{2}\beta\left[\cos(\gamma)+\sin(\gamma)\right]\vert V\rangle
\end {aligned}
\end{equation*}
Actually, we perform the measurement with the angle $\phi=\pi/8$.
In order to simplify the experimental setup only the transmitted path of the polarizing beam splitter is used.
 The measuring of the reflective path (the dashed lines in Fig.\ 1) can be simulated by setting $\phi=-\pi/8$. The corresponding (subnormalized) density matrices are
\begin{equation*}
\mc I^\gamma_\pm(\kb{\Psi_{\rm in}}{\Psi_{\rm in}}):=\kb{\Psi^\pm_9}{\Psi^\pm_9}=\frac12
\begin{pmatrix}
\alpha\overline\alpha[\cos(\gamma)\pm\sin(\gamma)]^2 & \alpha\overline\beta[\cos^2(\gamma)-\sin^2(\gamma)]\\
\beta\overline\alpha[\cos^2(\gamma)-\sin^2(\gamma)] & \beta\overline\beta[\cos(\gamma)\mp\sin(\gamma)]^2
\end{pmatrix}
\end{equation*}
Hence, we have the quantum operations 
\begin{eqnarray*}
\mc I^\gamma_\pm(\varrho) 
&=&\frac12\begin{pmatrix}
[\cos(\gamma)\pm\sin(\gamma)]^2 & \cos^2(\gamma)-\sin^2(\gamma) \\
 \cos^2(\gamma)-\sin^2(\gamma) & [\cos(\gamma)\mp\sin(\gamma)]^2
\end{pmatrix}\star\varrho \\
&=&\frac12\begin{pmatrix}
1\pm\sin(2\gamma) & \cos(2\gamma) \\
\cos(2\gamma) & 1\mp\sin(2\gamma)
\end{pmatrix}\star\varrho =K_{\pm\gamma}\varrho K_{\pm\gamma}
\end{eqnarray*}
where $\varrho$ is any initial state (density matrix), $\star$ is the entrywise (i.e.\ Schur) multiplication in the basis $\{\ket H,\ket V\}$, and
$$
K_{\pm\gamma}:=\frac{1}{\sqrt{2}}[\cos{(\gamma)}\id\pm\sin{(\gamma)}\sigma_z]=\frac{1}{\sqrt{2}}
\begin{pmatrix}
\cos(\gamma)\pm\sin(\gamma) & 0 \\
0 & \cos(\gamma)\mp\sin(\gamma) 
\end{pmatrix}
$$
The sum map $\Phi^\gamma:=\mc I^\gamma_++\mc I^\gamma_-$ is easily seen to be trace preserving, hence a trace-preserving completely positive (CPTP) map or a quantum channel which is of the Schur form. Indeed,
$$
\Phi^\gamma(\varrho)=\mc I^\gamma_+(\varrho)+\mc I^\gamma_-(\varrho)=\begin{pmatrix}
1 & \cos(2\gamma) \\
\cos(2\gamma) & 1
\end{pmatrix}\star\varrho.
$$
Hence, $\mc I^\gamma:=(\mc I^\gamma_+,\mc I^\gamma_-)$ is a (Schr\"odinger) instrument. 
Note that the dual (Heisenberg) instrument $\mc I^{\gamma\,*}$ is mathematically of the same form as $\mc I^\gamma$ (the dual is defined via $\tr[\mc I^{\gamma\,*}_\pm(M)\varrho]:=\tr[M\mc I^\gamma_\pm(\varrho)]$ for all operators (matrices) $M$ and $\varrho$).  
As a conclusion, we have an instrument $\mc I^\gamma$ which measures the POVM $\ms A^\gamma=(A^\gamma_+,A^\gamma_-)$ where 
$$
A^\gamma_\pm=\mc I^{\gamma\,*}_\pm(\id)=
\frac12\begin{pmatrix}
1\pm\sin(2\gamma) & \cos(2\gamma) \\
\cos(2\gamma) & 1\mp\sin(2\gamma)
\end{pmatrix}\star\id
=\frac12\big[\id\pm\sin{(2\gamma)}\sigma_z\big].
$$
 This POVM is a noisy version (mixture with white noise) of the sharp POVM $\ms Z=(Z_+,Z_-)$ where $Z_\pm=(1/2)(\id\pm\sigma_z)$. Indeed, $\lambda Z_{\pm}+(1-\lambda)\frac12\id=A^\gamma_\pm$ where $\lambda=\sin{(2\gamma)}$.

  It is interesting to note that $K_{\pm\gamma}^2=A^\gamma_\pm$ and, in the region $\gamma\in[0,\pi/4]$, $K_{\pm\gamma}$ are positive semidefinite. Thus, for this parameter range, $\mc I^\gamma_\pm(\varrho)=(A^\gamma_\pm)^{1/2}\varrho(A^\gamma_\pm)^{1/2}$ for all input states $\varrho$, i.e.,\ $\mc I$ represents the L\"{u}ders measurement of $\ms A^\gamma$ whenever $\gamma\in[0,\pi/4]$.
  
\begin{remark}
 \textit{Realised instrument.} Our measurement scenario produces the least disturbing implementation of noisy $Z$, i.e. the L\"uders instrument without the outcome dependent quantum noise channels.
\end{remark}

\noindent $\boxed{10}$ 
After this particular measurement we perform a full tomography with rank-1 projections 
 $$
 \kb\fii\fii=\begin{pmatrix}
 |c|^2 & c\overline d \\
 \overline c d & |d|^2
 \end{pmatrix}
 $$
 where $\fii=c\ket H+d\ket V$, $c,\,d\in\C$, $|c|^2+|d|^2=1$. Now
 $$
 \mc I^{\gamma\,*}_\pm( \kb\fii\fii)=\frac12\begin{pmatrix}
1\pm\sin(2\gamma) & \cos(2\gamma) \\
\cos(2\gamma) & 1\mp\sin(2\gamma)
\end{pmatrix}\star\begin{pmatrix}
 |c|^2 & c\overline d \\
 \overline c d & |d|^2
 \end{pmatrix}
 $$
 and the measurement probabilities are
 \begin{eqnarray*}
 \<\fii|\mc I^\gamma_\pm(\varrho)\fii\>&=&\tr[\mc I^{\gamma\,*}_\pm( \kb\fii\fii)\varrho]=\<\Psi_{\rm in}|\mc I^{\gamma\,*}_\pm( \kb\fii\fii)\Psi_{\rm in}\> \\
 &=&
 \frac{1}{2}\left[
 |\alpha|^2|c|^2+|\beta|^2|d|^2+(\alpha\overline\beta\overline c d+\overline\alpha\beta c\overline d)\cos(2\gamma)\pm
\left( |\alpha|^2|c|^2-|\beta|^2|d|^2\right)\sin(2\gamma) \right]
\end{eqnarray*}
 where the input state $\varrho=\kb{\Psi_{\rm in}}{\Psi_{\rm in}}$, $\Psi_{\rm in}=\alpha\ket H+\beta\ket V$.
 We get the following tomography table:
\begin{center}
 \begin{tabular}{ c | c | c }
 Projection $\kb\fii\fii$ & $\fii$ & Probability $\<\fii|\mc I^\gamma_\pm(\varrho)\fii\>=\<\Psi_{\rm in}|\mc I^{\gamma\,*}_\pm(\kb\fii\fii)\Psi_{\rm in}\>$ \\
 \hline 
  $X_+=\frac12(\id+\sigma_x)$ &
 $\frac{1}{\sqrt2}(\ket H+\ket V)$ & $ \frac{1}{2}\left[
 \frac{1}{2}+{\rm Re}(\alpha\overline\beta)\cos(2\gamma)\pm\frac12\left( |\alpha|^2-|\beta|^2\right)\sin(2\gamma) \right] $\\
 $X_-=\frac12(\id-\sigma_x)$ &
 $\frac{1}{\sqrt2}(\ket H-\ket V)$ & $ \frac{1}{2}\left[
 \frac{1}{2}-{\rm Re}(\alpha\overline\beta)\cos(2\gamma)\pm\frac12\left( |\alpha|^2-|\beta|^2\right)\sin(2\gamma) \right] $\\
  $Y_+=\frac12(\id+\sigma_y)$ &
 $\frac{1}{\sqrt2}(\ket H+i\ket V)$ & $ \frac{1}{2}\left[
 \frac{1}{2}-{\rm Im}(\alpha\overline\beta)\cos(2\gamma)\pm\frac12\left( |\alpha|^2-|\beta|^2\right)\sin(2\gamma) \right] $\\
 $Y_-=\frac12(\id-\sigma_y)$ &
 $\frac{1}{\sqrt2}(\ket H-i\ket V)$ & $ \frac{1}{2}\left[
 \frac{1}{2}+{\rm Im}(\alpha\overline\beta)\cos(2\gamma)\pm\frac12\left( |\alpha|^2-|\beta|^2\right)\sin(2\gamma) \right] $\\
$Z_+=\frac12(\id+\sigma_z)$ &
 $\ket H$ & $ \frac12|\alpha|^2\left[1\pm \sin(2\gamma) \right] $\\
 $Z_-=\frac12(\id-\sigma_z)$ &
$\ket V$ & $ \frac{1}{2} |\beta|^2\left[1\mp\sin(2\gamma) \right] $
 \end{tabular}
 \end{center}
 From this table, one gets the sequential probabilities of the first $\ms A^\gamma$-measurement followed by a sharp POVM, say, $(X_+,X_-)$ (see the first two rows).
Next we present theoretical probabilities (e.g.\ in the row $X_+$) and the corresponding experimental values. One sees that the measured probabilities are quite close to their theoretical values. The statistical error of the measured probabilities are $\pm$ 2\%. 
 
 \section{Case $\gamma=0$}
\begin{center}
 \begin{tabular}{ l | c | c | c | c | c | c | c | c | c | c}
 $\Psi_{\rm in}=$ & \multicolumn{2}{| c |}{$\ket H$}  &  \multicolumn{2}{| c |}{$\ket V$}  & \multicolumn{2}{| c |}{$\frac{1}{\sqrt2}(\ket H+\ket V)$}  & 
 \multicolumn{2}{| c |}{$\frac{1}{\sqrt2}(\ket H-\ket V)$} & \multicolumn{2}{| c |}{$\cos\frac\pi8\ket H+\sin\frac\pi8\ket V$}  \\ \hline
 $\mathcal I^0_\pm=$ & + & -- & + & -- & + & -- & + & -- &  +  & -- \\ \hline
$X_+$ {\rm(theory)}&  1/4 & 1/4 & 1/4  & 1/4 & 1/2 & 1/2 & 0 & 0 & 0.427 & 0.427   \\ 
$X_+$ {\rm(experiment)} & 	0.254	&0.255	&	0.248	&0.261	&	0.469&	0.479	&	0.017	&0.016		&0.420	&0.423 \\
$X_-$ {\rm(theory)}& 1/4 & 1/4 & 1/4 & 1/4 & 0 & 0 & 1/2 & 1/2 & 0.073 & 0.073 \\
$X_-$ {\rm(experiment)} &	0.246	&0.246	&	0.245	&0.246		&0.024	&0.028		&0.479	&0.488		&0.082	&0.076\\
$Y_+$ {\rm(theory)}& 1/4 & 1/4 & 1/4 & 1/4 & 1/4 & 1/4 & 1/4 & 1/4 & 1/4 & 1/4  \\
$Y_+$ {\rm(experiment)} 	&0.252	&0.251	&	0.237&	0.254	&	0.237	&0.254	&	0.250	&0.261	&	0.261	&0.249\\
$Y_-$ {\rm(theory)}& 1/4 & 1/4 & 1/4 & 1/4 & 1/4 & 1/4 & 1/4 & 1/4 &  1/4 & 1/4 \\
$Y_-$ {\rm(experiment)} 	&0.250	&0.247	&	0.248	&0.262		&0.248&	0.262	&	0.247	&0.242		&0.247	&0.243\\
$Z_+$ {\rm(theory)}& 1/2 & 1/2 & 0 & 0 & 1/4 & 1/4 & 1/4 & 1/4 &  0.427 & 0.427   \\ 
$Z_+$ {\rm(experiment)} 	&0.494	&0.497	&	0.008	&0.009	&	0.252	&0.249		&0.246	&0.245		&0.431	&0.423\\
$Z_-$ {\rm(theory)}& 0  & 0 & 1/2 & 1/2 & 1/4 & 1/4 & 1/4 & 1/4 &  0.073 & 0.073 \\
$Z_-$ {\rm(experiment)} &	0.004&	0.004	&	0.480	&0.503	&	0.243	&0.256	&	0.246&	0.262	&	0.070	& 0.076 \\
 \end{tabular}
 \end{center}

  \section{Case $\gamma=\pi/8$}
\begin{center}
 \begin{tabular}{ l | c | c | c | c | c | c | c | c | c | c}
 $\Psi_{\rm in}=$ & \multicolumn{2}{| c |}{$\ket H$}  &  \multicolumn{2}{| c |}{$\ket V$}  & \multicolumn{2}{| c |}{$\frac{1}{\sqrt2}(\ket H+\ket V)$}  & 
 \multicolumn{2}{| c |}{$\frac{1}{\sqrt2}(\ket H-\ket V)$} & \multicolumn{2}{| c |}{$\cos\frac\pi8\ket H+\sin\frac\pi8\ket V$}  \\ \hline
 $\mathcal I^{\pi/8}_\pm=$ & + & -- & + & -- & + & -- & + & -- &  +  & -- \\ \hline
$X_+$ {\rm(theory)} & 0.427  & 0.073 & 0.073  & 0.427 & 0.427 & 0.427 &  0.073 & 0.073  & 1/2 & 1/4  \\ 
$X_+$ {\rm(experiment)} 	& 0.417	& 0.079		& 0.087	& 0.414		& 0.414	& 0.426		& 0.075	& 0.089		& 0.487	& 0.254\\ 
$X_-$ {\rm(theory)} &   0.427  & 0.073  &  0.073 & 0.427 & 0.073 & 0.073 &  0.427 & 0.427  & 1/4 & 0  \\ 
$X_-$ {\rm(experiment)} 	& 0.414	& 0.090		& 0.077	& 0.422		& 0.075	& 0.085		& 0.425	& 0.411		& 0.243	& 0.016\\ 
$Y_+$ {\rm(theory)} &   0.427  & 0.073  &  0.073 & 0.427 & 1/4 & 1/4 &  1/4 & 1/4  & 3/8 = 0.375 & 1/8 = 0.125 \\ 
$Y_+$ {\rm(experiment)} 	& 0.426	& 0.099		& 0.066	& 0.423		& 0.258	& 0.247		& 0.246	& 0.255		& 0.381	& 0.126\\ 
$Y_-$ {\rm(theory)} &   0.427  & 0.073  & 0.073  & 0.427 & 1/4 & 1/4 &  1/4 & 1/4  & 3/8 & 1/8  \\ 
$Y_-$ {\rm(experiment)} 	& 0.415	& 0.060		& 0.077	& 0.434		& 0.249	& 0.245		& 0.237	& 0.262		& 0.373	& 0.120\\ 
$Z_+$ {\rm(theory)} & 0.854  & 0.146 &  0 & 0 & 0.427 & 0.073 &  0.427 & 0.073  & 0.729 & 1/8 \\ 
$Z_+$ {\rm(experiment)} 	& 0.844	& 0.143		& 0.006	& 0.007		& 0.414	& 0.090		& 0.425	& 0.070		& 0.710	& 0.131\\ 
$Z_-$ {\rm(theory)} & 0  & 0 & 0.146  & 0.854  & 0.073 & 0.427 &  0.073 & 0.427  & 0.021 & 1/8  \\ 
$Z_-$ {\rm(experiment)} 	& 0.006	& 0.006		& 0.147	& 0.840		& 0.063	& 0.433		& 0.092	& 0.413		& 0.031	& 0.128\\ 
 \end{tabular}
 \end{center}
 
  \section{Case $\gamma=\pi/4$}
\begin{center}
 \begin{tabular}{ l | c | c | c | c | c | c | c | c | c | c}
 $\Psi_{\rm in}=$ & \multicolumn{2}{| c |}{$\ket H$}  &  \multicolumn{2}{| c |}{$\ket V$}  & \multicolumn{2}{| c |}{$\frac{1}{\sqrt2}(\ket H+\ket V)$}  & 
 \multicolumn{2}{| c |}{$\frac{1}{\sqrt2}(\ket H-\ket V)$} & \multicolumn{2}{| c |}{$\cos\frac\pi8\ket H+\sin\frac\pi8\ket V$}  \\ \hline
 $\mathcal I^{\pi/4}_\pm=$ & + & -- & + & -- & + & -- & + & -- &  +  & -- \\ \hline
$X_+$ {\rm(theory)} &1/2 & 0  & 0 & 1/2 & 1/4 & 1/4 & 1/4 &  1/4 & 0.427 & 0.073    \\ 
$X_+$ {\rm(experiment)}	& 0.483	& 0.019		& 0.025	& 0.470		& 0.248	& 0.257		& 0.245	& 0.243		& 0.419	& 0.082\\
$X_-$ {\rm(theory)}  &1/2 & 0  & 0 & 1/2 & 1/4 & 1/4 & 1/4 &  1/4 & 0.427  &  0.073   \\ 
$X_-$ {\rm(experiment)}	& 0.476	& 0.023		& 0.023	& 0.481		& 0.255	& 0.241		& 0.257	& 0.256		& 0.417	& 0.082\\
$Y_+$ {\rm(theory)} &1/2 & 0  & 0 & 1/2 & 1/4 & 1/4 & 1/4 &  1/4 & 0.427  &  0.073   \\ 
$Y_+$ {\rm(experiment)}	& 0.478	& 0.026		& 0.023	& 0.479		& 0.235	& 0.263		& 0.257	& 0.252		& 0.416	& 0.070\\
$Y_-$ {\rm(theory)} &1/2 & 0 & 0 & 1/2 & 1/4 & 1/4 & 1/4 & 1/4  & 0.427  &   0.073  \\ 
$Y_-$ {\rm(experiment)}	& 0.489	& 0.008		& 0.010	& 0.488		& 0.255	& 0.247		& 0.250	& 0.241		& 0.425	& 0.089\\
$Z_+$ {\rm(theory)} &1 & 0  & 0 & 0 & 1/2 & 0 & 1/2 & 0  &  0.854 &  0   \\ 
$Z_+$ {\rm(experiment)}	& 0.968	& 0.022		& 0.004	& 0.006		& 0.452	& 0.017		& 0.492	& 0.006		& 0.842	& 0.020\\
$Z_-$ {\rm(theory)} &0 & 0  & 0 & 1 & 0 & 1/2 & 0 & 1/2  &  0  &  0.146   \\ 
$Z_-$ {\rm(experiment)}	& 0.005	& 0.004		& 0.027	& 0.963		& 0.017	& 0.513		& 0.012	& 0.489		& 0.010	& 0.127\\
 \end{tabular}
 \end{center}
 
 Note that $\cos\frac\pi8=\frac1{\sqrt2}\sqrt{1+\frac1{\sqrt2}}$ and $\sin\frac\pi8=\frac1{\sqrt2}\sqrt{1-\frac1{\sqrt2}}$. In the above  tables we have used the approximations
$\frac14\left(1+\frac1{\sqrt2}\right)\approx 0.427$,
$\frac14\left(1-\frac1{\sqrt2}\right)\approx 0.073$,
 $\frac12\left(1+\frac1{\sqrt2}\right)\approx 0.854$, 
$\frac12\left(1-\frac1{\sqrt2}\right)\approx 0.146$,
$\frac14\left(\frac32+\sqrt2\right)\approx0.729$, and
$\frac14\left(\frac32-\sqrt2\right)\approx0.021$.

 We have the following interesting cases:
 \begin{enumerate}
 \item \textit{No first measurement.} Here the parameter $\gamma=0$. Now $A^0_\pm=\frac12\id$ is a trivial (white noise) POVM, $\mc I^0_\pm(\varrho)=\frac12\varrho$, and $\Phi^0(\varrho)=\varrho$ is the identity channel. This corresponds to the case where there is no first measurement.  Moreover, the measurement probabilities are 
 $$\<\Psi_{\rm in}|\mc I^{0\,*}_\pm( \kb\fii\fii)\Psi_{\rm in}\>=\frac12\big|\<\fii|\Psi_{\rm in}\>\big|^2.$$
 Especially, if $\alpha$ and $\beta$ are real numbers, then the final $\ms Y$ measurement (that is, $\fii=(\ket H\pm i\ket V)/\sqrt2$) gives the totally random probabilities $\frac14$. Moreover, one can measure the random noise POVM:
 $$
 \<\Psi_{\rm in}|A^0_\pm\Psi_{\rm in}\>=\<\Psi_{\rm in}|\mc I^{0\,*}_\pm(\id)\Psi_{\rm in}\>=\<\Psi_{\rm in}|\mc I^{0\,*}_\pm(Y_+)\Psi_{\rm in}\>+\<\Psi_{\rm in}|\mc I^{0\,*}_\pm(Y_-)\Psi_{\rm in}\>=\frac12.
 $$
 
\item \textit{The first measurement is sharp $\ms Z$.} Here we have $\gamma=\pi/4$ and one gets  
 $\mc I^{\pi/4}_\pm(\varrho)=Z_\pm\varrho Z_\pm$ and $\Phi^{\pi/4}(\varrho)=Z_+\varrho Z_++Z_-\varrho Z_-$, i.e.,\ the total channel of the measurement is the decohering channel in the basis $\{\ket H,\ket V\}$  and, indeed, allows the measurement of the sharp POVM $\ms A^{\pi/4}=\ms Z$.
 
\item \textit{Macrorealistic case: the final measurement is sharp $\ms Z$.} Here $\ms Z=(Z_+,Z_-)$, $Z_+=\kb HH$, $Z_-=\kb VV$, which generates the joint POVM $\ms G^\gamma=(G^\gamma_{++},G^\gamma_{+-},G^\gamma_{-+},G^\gamma_{--})$ where
 $$
G^\gamma_{a,b}:=\mc I^{\gamma\,*}_a(Z_b)=\frac{1}{2}[1+ab\sin{(2\gamma)}]Z_b,\qquad a,b=\pm1,
$$
which is a smearing of $\ms Z$ with the conditional probabilites $\frac{1}{2}[1+ab\sin{(2\gamma)}]$. {\it For all $\gamma$}, the margins are $\ms A^\gamma$ (a noisy $\ms Z$) and the sharp $\ms Z$  since $\Phi^{\gamma\,*}(Z_\pm)=\mc I^{\gamma\,*}_+(Z_\pm)+\mc I^{\gamma\,*}_-(Z_\pm)=Z_\pm$. From items (1) and (2) we see that one can measure sharp $\ms Z$ (the case $\gamma=\pi/4$) or any noisy $\ms A^\gamma$ first and then $\ms Z$ to get the same second margin $\ms Z$, or `one needs not perform any measurement' at first (the case $\gamma=0$). Indeed, 
$$
\<\Psi_{\rm in}|\mc I^{\gamma\,*}_+(Z_+)\Psi_{\rm in}\>+\<\Psi_{\rm in}|\mc I^{\gamma\,*}_-(Z_+)\Psi_{\rm in}\>
=\big|\<H|\Psi_{\rm in}\>\big|^2=|\alpha|^2
$$
and experimental results of these cases (when $\gamma=0,\;\pi/8,\;\pi/4$) with theoretical predictions for the probabilities $\big|\<H|\Psi_{\rm in}\>\big|^2$ are shown in the Fig.\ 2 of the Letter.
 
\item \textit{Retrieving scenario: the final measurement is sharp $\ms X$.} Here $\ms X=(X_+,X_-)$ with the joint POVM 
  $\ms G^\gamma=(G^\gamma_{++},G^\gamma_{+-},G^\gamma_{-+},G^\gamma_{--})$ where
$$\boxed{
G^\gamma_{a,b}:=\mc I^{\gamma\,*}_a(X_b)=
\frac{1}{4}\big[\id+a\sin{(2\gamma)}\sigma_z+b\cos{(2\gamma)}\sigma_x\big].
}$$
The first margin of $\ms G^\gamma$ is naturally the POVM $\ms A^\gamma$ measured first and the second margin is immediately found to be a noisy $\ms X$-POVM $\ms B^\gamma=(B^\gamma_+,B^\gamma_-)$ where $B^\gamma_\pm=\Phi^{\gamma\,*}(X_\pm)=\frac12\big[\id\pm\cos{(2\gamma)}\sigma_x\big]$. When $\gamma=0$ we get
$\ms B^0=\ms X$. 
Note that if we measure $\ms Y$ instead of $\ms X$ we get similar joint POVM $\ms G^\gamma$ where $\sigma_x$ is replaced by $\sigma_y$.

\item \textit{Retrieved scenario: final measurement is noisy $\ms X$.} Now we measure $\ms X^\lambda=(X^\lambda_+,X^\lambda_-)$ where 
$$
X^\lambda_{\pm}=\lambda X_{\pm}+(1-\lambda)\frac12\id=\frac12\left(\id\pm \lambda\sigma_x\right),\qquad \lambda\in[0,1]
$$
and there is no first measurement.
This can be measured by mixing the measurement (marginal) probabilities of the sharp $\ms X$ and the trivial POVM $\ms A^0$ in the case $\gamma=0$ [items (1), (4)]:
\begin{eqnarray*}
&&\<\Psi_{\rm in}|X^\lambda_\pm\Psi_{\rm in}\>=
\lambda\<\Psi_{\rm in}|X_\pm\Psi_{\rm in}\>+(1-\lambda)\<\Psi_{\rm in}|A^0_\pm\Psi_{\rm in}\> \\
&&=
\lambda\big[\<\Psi_{\rm in}|\mc I^{0\,*}_+(X_\pm)\Psi_{\rm in}\>+\<\Psi_{\rm in}|\mc I^{0\,*}_-(X_\pm)\Psi_{\rm in}\>\big]+(1-\lambda)\big[\<\Psi_{\rm in}|\mc I^{0\,*}_\pm(Y_+)\Psi_{\rm in}\>+\<\Psi_{\rm in}|\mc I^{0\,*}_\pm(Y_-)\Psi_{\rm in}\>\big].
\end{eqnarray*}
This should give the same results for $\ms B^\gamma$ of item (4) when $\lambda$ is chosen to be $\cos{(2\gamma)}$.
\end{enumerate}

\subsection{Optimal measurements and states}

From here on, we consider item (4) and the POVM $\ms G^\gamma$ which
is a joint measurement for the noisy $\ms Z$-POVM $\ms A^\gamma$ and the noisy 
$\ms X$-POVM $\ms B^\gamma$ since
$$
A^\gamma_\pm
=\frac12\big[\id\pm\sin{(2\gamma)}\sigma_z\big]=G^\gamma_{\pm,+}+G^\gamma_{\pm,-},\qquad
 B^\gamma_\pm=\frac12\big[\id\pm\cos{(2\gamma)}\sigma_x\big]=G^\gamma_{+,\pm}+G^\gamma_{-,\pm}.
$$
Next we show that the choice $\gamma=\pi/8$ (where $\sin{(2\gamma)}=\cos{(2\gamma)}=1/\sqrt2$) gives the optimal approximate joint measurement
$\ms G^{\pi/8}$ for $\ms Z$ and $\ms X$.

For any binary probability distribution $\{+1\}\mapsto p$, $\{-1\}\mapsto 1-p$, which we denote by $\mathsf p=(p,1-p)$,
the mean value $\overline{\mathsf p}$ is 
$p-(1-p)=2p-1$ and the variance ${\rm Var}(\mathsf p)$ is $(+1)^2p+(-1)^2(1-p)-\overline{\mathsf p}^2=1-\overline{\mathsf p}^2=4p(1-p)$. In our case, 
$\varrho=\frac12(\id+r_x\sigma_x+r_y\sigma_y+r_z\sigma_z)$ is a state, i.e., the real numbers $r_{\cdots}$ satisfy  $r_x^2+r_y^2+r_z^2\le 1$, and we get the following probabilities:
\begin{eqnarray*}
\mu_{ab}&:=&\tr[\varrho\, G^\gamma_{a,b}]=\tr[\varrho\,\mc I^{\gamma\,*}_a(X_b)]=
\frac{1}{4}\big[\id+a\sin{(2\gamma)}r_z+b\cos{(2\gamma)}r_x\big],\\
p&:=&\mu_{++}+\mu_{+-}
=\tr[A_+^\gamma\varrho]=\frac12[1+r_z\sin(2\gamma)], \\
q&:=&\mu_{++}+\mu_{-+}
=\tr[B_+^\gamma\varrho]=\frac12[1+r_x\cos(2\gamma)], \\
\tilde p&:=&\tr[Z_+\varrho]=\frac12(1+r_z), \\
\tilde q&:=&\tr[X_+\varrho]=\frac12(1+r_x).
\end{eqnarray*}
Recall from page 333 of Ref.~\cite{busch16} that the square of the Wasserstein-2 distance $\Delta_2(\tilde{\mathsf q},\mathsf q)$ for 2-valued probability distributions $\tilde{\mathsf q}$ and $\mathsf q$ is $\Delta_2(\tilde{\mathsf q},\mathsf q)^2=4|\tilde{q}-q|=2[1-\cos(2\gamma)]|r_x|$ whose maximum $\Delta(\ms B^\gamma,\ms X)^2=2[1-\cos(2\gamma)]$ is attained when $r_x=\pm1$. Similarly, $\Delta_2(\tilde{\mathsf p},\mathsf p)^2=4|\tilde{p}-p|=2[1-\sin(2\gamma)]|r_z|\le 2[1-\sin(2\gamma)]=\Delta(\ms A^\gamma,\ms Z)^2$ so 
$$
\Delta(\ms A^\gamma,\ms Z)^2+\Delta(\ms B^\gamma,\ms X)^2=2[2-\sin(2\gamma)-\cos(2\gamma)]
\ge2\big(2-\sqrt{2}\big)
$$
where the minimum is at the point $\gamma=\pi/8$. Actually, according to Ref.~\cite{BLW2014}, the minimum value $2(2-\sqrt{2})$ for the uncertainty sum $\Delta(\ms A,\ms Z)^2+\Delta(\ms B,\ms X)^2$ is reached for $\ms A=\ms A^{\pi/8}$ and $\ms B=\ms B^{\pi/8}$ where $\Delta$ is the (above) {\it worst-case} Wasserstein-2 distance for POVMs (and $\ms A$ and $\ms B$ are any jointly measurable binary POVMs). 
\begin{remark}
\textit{Optimal measurements.} The measurement setting for $\gamma=\pi/8$ realizes the optimal approximate joint measurement of $\ms Z$ and $\ms X$. We note that this is reached by measuring the L\"uders instrument of noisy $Z$ (without noise channels).
\end{remark}

Note that, when $\gamma=\pi/8$, $$\Delta_2(\tilde{\mathsf q},\mathsf q)^2+\Delta_2(\tilde{\mathsf p},\mathsf p)^2=(2-\sqrt2)(|r_x|+|r_z|)\le 2(\sqrt2-1)\approx0.828$$ where the maximum is obtained when $|r_x|=|r_z|=1/\sqrt2$, e.g.\ $\Psi_{\rm in}=\cos\frac\pi8\ket H+\sin\frac\pi8\ket V$. The minimum of the uncertainty sum is 0 when $r_y=\pm1$, i.e.\ $\Psi_{\rm in}=\frac{1}{\sqrt2}(\ket H\pm i\ket V)$ (In the worst case scenario, the corresponding maximum sum is $\sup_\varrho\Delta_2(\tilde{\mathsf q},\mathsf q)^2+\sup_\varrho\Delta_2(\tilde{\mathsf p},\mathsf p)^2=2(2-\sqrt2)$.) 
In theory, we have
 \begin{center}
 \begin{tabular}{ c | c | c | c | c | c }
 $\Psi_{\rm in}=$ & {$\ket H$}  &  {$\ket V$}  & {$\frac{1}{\sqrt2}(\ket H+\ket V)$}  & 
{$\frac{1}{\sqrt2}(\ket H-\ket V)$} & {$\cos\frac\pi8\ket H+\sin\frac\pi8\ket V$}  \\ \hline
$\Delta_2(\tilde{\mathsf q},\mathsf q)^2+\Delta_2(\tilde{\mathsf p},\mathsf p)^2=$ & 0.586 & 0.586 & 0.586 & 0.586 & 0.828
 \end{tabular}
 \end{center}
 and the experimental values are
 \begin{center}
 \begin{tabular}{ c | c | c | c | c | c }
 $\Psi_{\rm in}=$ & {$\ket H$}  &  {$\ket V$}  & {$\frac{1}{\sqrt2}(\ket H+\ket V)$}  & 
{$\frac{1}{\sqrt2}(\ket H-\ket V)$} & {$\cos\frac\pi8\ket H+\sin\frac\pi8\ket V$}  \\ \hline
$\Delta_2(\tilde{\mathsf q},\mathsf q)^2+\Delta_2(\tilde{\mathsf p},\mathsf p)^2=$ & 0.692 & 0.620 & 0.480 & 0.560 & 0.904 
 \end{tabular}
 \end{center}
Next we study the uncertainty sum $S^\gamma_\varrho$ of the variances of $\ms A^\gamma$ and $\ms B^\gamma$ in a state $\varrho$ and find the minimum uncertainty states (which minimize this sum):
\begin{eqnarray*}
{\rm Var}(\ms A^\gamma,\varrho)&=&4p(1-p)=1-r_z^2\sin^2(2\gamma), \\
{\rm Var}(\ms B^\gamma,\varrho)&=&4q(1-q)=1-r_x^2\cos^2(2\gamma), \\
S^\gamma_\varrho&:=&
{\rm Var}(\ms A^\gamma,\varrho)+{\rm Var}(\ms B^\gamma,\varrho)=2-r_x^2\cos^2(2\gamma)-r_z^2\sin^2(2\gamma)\\
&\ge& 2-r_x^2\cos^2(2\gamma)-(1-r_x^2)\sin^2(2\gamma) 
=2-\sin^2(2\gamma)-r_x^2[1-2\sin^2(2\gamma)]\\
&=&2-\sin^2(2\gamma)-r_x^2\cos(4\gamma) \\
&\ge&
\begin{cases}
2-\cos^2(2\gamma), &
0\le \gamma\le \frac{\pi}{8}, \\
2-\sin^2(2\gamma), &
\frac{\pi}{8}\le \gamma\le \frac{\pi}{4},
\end{cases}
\end{eqnarray*}
where the lower bounds are reached in the cases $r_x=\pm1$, i.e.\ $\Psi_{\rm in}=2^{-1/2}(\ket H\pm\ket V)$ (when $\gamma< \frac{\pi}{8}$) and $r_z=\pm1$, i.e.\ $\Psi_{\rm in}$ is either $\ket H$ or $\ket V$ (when $\gamma> \frac{\pi}{8}$).

\begin{remark}
\textit{Minimum uncertainty states.} When $\gamma=\frac{\pi}{8}$ then the minimum uncertainty $\frac32$ is reached for all pure states with $r_y=0$, i.e.\ $\Psi_{\rm in}$ is any {\it real} superposition of the basis vectors.
Hence, it is reasonable to prepare only these kinds of pure states.
\end{remark}

Let us first consider the extreme cases $\gamma=\pi/4$ or $0$. Now $\ms A^{\pi/4}=\ms Z$ and $B^{\pi/4}_\pm=\frac12\id$ whereas  $A^0_\pm=\frac12\id$ and $\ms B^0=\ms X$. In both cases, the minimum sum of variances is 1 since the variance of the trivial POVM is $4\cdot\frac12\big(1-\frac12\big)=1$ (in any state) and the variance of a sharp POVM is 0 in its eigenstate. In the case $\gamma=\pi/8$, we have listed the variances in the following table for the states that we have used in the experiment:
\begin{center}
 \begin{tabular}{ l | c | c | c | c | c | c | c}
$\Psi_{\rm in}=$ & {$\ket H$}  &  {$\ket V$}  & {$\frac{1}{\sqrt2}(\ket H+\ket V)$}  & 
{$\frac{1}{\sqrt2}(\ket H-\ket V)$} & {$\cos\frac\pi8\ket H+\sin\frac\pi8\ket V$}  \\ \hline
 $p$ theory & 0.854 & 0.146 & 1/2 & 1/2 & 3/4 \\
 $p$ experiment & 0.831 & 0.164 & 0.489 & 0.500 & 0.730 \\ \hline
 $q$ theory & 1/2 & 1/2 & 0.854 & 0.146 & 3/4 \\
 $q$ experiment & 0.496 & 0.501 & 0.840 & 0.164 & 0.741 \\ \hline
 ${\rm Var}(\ms A^{\pi/8},\varrho)$ th.& 1/2 & 1/2 & 1 & 1 & 3/4 \\
  ${\rm Var}(\ms A^{\pi/8},\varrho)$ exp.& 0.562 & 0.548 & 1.000 & 1.000 & 0.788 \\ \hline
  ${\rm Var}(\ms B^{\pi/8},\varrho)$ th.& 1 & 1 & 1/2 & 1/2 & 3/4 \\
  ${\rm Var}(\ms B^{\pi/8},\varrho)$ exp.& 1.000 & 1.000 & 0.538 & 0.548 & 0.768 \\ \hline
  $S^{\pi/8}_\varrho$ experiment & 1.562 & 1.548  & 1.538 & 1.548 & 1.556
 \end{tabular}
 \end{center}

Finally, we study correlation in POVM $\ms G^\gamma$. For two probability distributions $\mathsf p=(p,1-p)$ and $\mathsf q=(q,1-q)$ (with values $\pm1$) with a joint probability distribution $\mu=(\mu_{++},\mu_{+-},\mu_{-+},\mu_{--})$ (i.e.\ $\mu_{++}+\mu_{+-}=p$ and $\mu_{++}+\mu_{-+}=q$) the correlation coefficient is
$$
{\rm Corr}(\mu)=
\frac{\sum_{a,b=\pm1}ab\,\mu_{ab}-\overline{\mathsf p}\,\overline{\mathsf q}}{\sqrt{{\rm Var}(\mathsf p)}\sqrt{{\rm Var}(\mathsf q)}}
\in[-1,1].
$$
In our case, $\mu_{ab}=\frac{1}{4}\big[1+a\sin{(2\gamma)}r_z+b\cos{(2\gamma)}r_x\big]$ whence
$\sum_{a,b=\pm1}ab\,\mu_{ab}=0$ and
$$
{\rm Corr}(\ms G^\gamma,\varrho)=-\frac{r_z\sin(2\gamma)\,r_x\cos(2\gamma)}{\sqrt{1-r_z^2\sin^2(2\gamma)}\sqrt{1-r_x^2\cos^2(2\gamma)}}.
$$
Note that in the cases $\gamma=0$ and $\gamma=\pi/4$ the correlation coefficient is not defined for all states (e.g.\ when $\gamma=0$ and $r_x=\pm1$ the second variance vanishes).
We find that
$$
-\frac{\sin(4\gamma)}{2+\sin(4\gamma)}\le
{\rm Corr}(\ms G^\gamma,\varrho)
\le\frac{\sin(4\gamma)}{2+\sin(4\gamma)}
$$
as long as $\gamma\in[0,\pi/4]$. 

\begin{remark}
\textit{Maximally (anti-)correlated states.} In the case $\gamma=\pi/8$, the above limits ${\rm Corr}(\ms G^{\pi/8},\varrho)=\pm\frac13$ are reached with pure states $\varrho=\kb{\Psi_{\rm in}}{\Psi_{\rm in}}$ where
$$
\Psi_{\rm in}=\cos\frac\pi8\ket H\mp\sin\frac\pi8\ket V
$$
(with $r_x=\mp1/\sqrt2$, $r_y=0$ and $r_z=1/\sqrt2$).
\end{remark}

Next we calculate from experimental data the correlation of the joint probability measure of observable $\ms G^{\pi/8}$ in state $\cos\frac\pi8\ket H+\sin\frac\pi8\ket V$. 
Theory gives the maximal anticorrelation $-1/3$ and the experimental value is  
\begin{eqnarray*}
{\rm Corr}(\ms G^{\pi/8},\varrho)&=&
\frac{\mu_{++}+\mu_{--}-\mu_{+-}-\mu_{-+}-(2p-1)(2q-1)}{\sqrt{{\rm Var}(\ms A^{\pi/8},\varrho){\rm Var}(\ms B^{\pi/8},\varrho)}} \\
&\approx&\frac{0.487+0.016-0.254-0.243-(2\cdot 0.730-1)(2\cdot0.741-1)}{
\sqrt{0.788\cdot0.768}}\\
&\approx&-0.28.
\end{eqnarray*}

\subsection{The retrieving case}

Now we continue to study item (5) in the case $\lambda=1/{\sqrt2}\approx0.707$. 
We compare the probabilities of a direct measurement of the noisy $\ms X$-POVM 
$$
X^{1/\sqrt{2}}_{\pm}=\frac12\left(\id\pm \frac1{\sqrt2}\sigma_x\right)
$$
and $\ms B^{\pi/8}$ obtained from a sequential measurement of $\ms A^{\pi/8}$ and sharp $\ms X$. 
Theoretically, since $X^{1/\sqrt{2}}=\ms B^{\pi/8}$, the probabilities are equal.
We use the formula (the case $\gamma=0$)
\begin{eqnarray*}
&&\<\Psi_{\rm in}|X^{1/\sqrt{2}}_+\Psi_{\rm in}\>=\\
&&
 \frac1{\sqrt2}\big[\<\Psi_{\rm in}|\mc I^{0\,*}_+(X_+)\Psi_{\rm in}\>+\<\Psi_{\rm in}|\mc I^{0\,*}_-(X_+)\Psi_{\rm in}\>\big]+\left(1- \frac1{\sqrt2}\right)\big[\<\Psi_{\rm in}|\mc I^{0\,*}_+(Y_+)\Psi_{\rm in}\>+\<\Psi_{\rm in}|\mc I^{0\,*}_+(Y_-)\Psi_{\rm in}\>\big]
\end{eqnarray*}
and present the experimental results in the next table:
  \begin{center}
 \begin{tabular}{ l | c | c | c | c | c }
 $\Psi_{\rm in}=$ & {$\ket H$}  &  {$\ket V$}  & {$\frac{1}{\sqrt2}(\ket H+\ket V)$}  & 
{$\frac{1}{\sqrt2}(\ket H-\ket V)$} & {$\cos\frac\pi8\ket H+\sin\frac\pi8\ket V$}  \\ \hline
$\<\Psi_{\rm in}|X^{1/\sqrt{2}}_+\Psi_{\rm in}\>=$ &  0.507 & 0.502 & 0.812 & 0.169 & 0.745  \\ 
$\<\Psi_{\rm in}|B^{\pi/8}_+\Psi_{\rm in}\>=$  & 0.496 & 0.501 & 0.840 & 0.164 & 0.741 \\ \hline
{\rm Theoretical values} & 1/2 & 1/2 & 0.854 & 0.146 & 3/4 \\
 \end{tabular}
 \end{center}
(Recall that $\<\Psi_{\rm in}|B^{\pi/8}_+\Psi_{\rm in}\>=\<\Psi_{\rm in}|\mc I^{\pi/8\,*}_+(X_+)\Psi_{\rm in}\>+\<\Psi_{\rm in}|\mc I^{\pi/8\,*}_-(X_+)\Psi_{\rm in}\>$.) These values are plotted in Fig.\ 3 of the Letter.

\subsection{The no-retrieving case} Suppose one wants to retrieve the sharp $\ms X$. The possible measurement probabilities are of the form
 $\<\fii|\Phi^\gamma(\varrho)\fii\>$ whereas we want to have, say, $\tr[\varrho X_+]=\frac12(1+r_x)$.
 If we choose $\varrho$ to be $ \kb{\Psi_{\rm in}}{\Psi_{\rm in}}$, where 
 $\Psi_{\rm in}=\frac{1}{\sqrt2}(\ket H+\ket V)$ is an eigenstate of the projection $X_+$ ($r_x=1$),
 we have $\tr[\varrho X_+]=1$.
 But 
 $
 \<\fii|\Phi^\gamma(\kb{\Psi_{\rm in}}{\Psi_{\rm in}})\fii\>=1
 $ 
 holds exactly when $\fii=\Psi_{\rm in}$ (i.e.\ $\kb\fii\fii=X_+$) and $\cos(2\gamma)=1$, that is,
 $A^\gamma_\pm=\frac12\id$, $\Phi^\gamma(\varrho)\equiv\varrho$ (no first measurement), and $B^\gamma_\pm=X_\pm$.
 
In the case $\gamma=0$ the first measurement is trivial, $A^0_\pm
=\frac12\id$, and the total channel is the identity channel (no measurement) so the related $\gamma=0$ data can be used to obtain the `sharp' probabilities $\tilde q=\<\Psi_{\rm in}|X_+\Psi_{\rm in}\>=\frac12(1+r_x)$. This can be compared to the probabilites $q=\<\Psi_{\rm in}|B^{\pi/8}_+\Psi_{\rm in}\>=\frac12[1+2^{-1/2}r_x]$ of the smeared $B^{\pi/8}_{\pm}=\frac12(\id\pm2^{-1/2}\sigma_x)$ in the same states. The square of the Wasserstein-2 distance is $\Delta_2(\tilde{\mathsf q},\mathsf q)^2=4|\tilde{q}-q|=(2-\sqrt2)|r_x|\le2-\sqrt2\approx0.586$ (where the maximum is given by states $2^{-1/2}(\ket H\pm \ket V)$).
 Theoretical values are:
 \begin{center}
 \begin{tabular}{ l | c | c | c | c | c }
 $\Psi_{\rm in}=$ & {$\ket H$}  &  {$\ket V$}  & {$\frac{1}{\sqrt2}(\ket H+\ket V)$}  & 
{$\frac{1}{\sqrt2}(\ket H-\ket V)$} & {$\cos\frac\pi8\ket H+\sin\frac\pi8\ket V$}  \\ \hline
$\tilde q=\<\Psi_{\rm in}|X_+\Psi_{\rm in}\>=$            & 1/2 & 1/2 & 1        & 0        & 0.854   \\
$q=\<\Psi_{\rm in}|B^{\pi/8}_+\Psi_{\rm in}\>=$ & 1/2 & 1/2 & 0.854 & 0.146 & 3/4 \\ \hline
$4|\tilde{q}-q|=\Delta_2(\tilde{\mathsf q},\mathsf q)^2=$ & 0 & 0 & 0.586 & 0.586 & 0.414 \\
 \end{tabular}
 \end{center}
 Measured values are:
 \begin{center}
 \begin{tabular}{ l | c | c | c | c | c }
 $\Psi_{\rm in}=$ & {$\ket H$}  &  {$\ket V$}  & {$\frac{1}{\sqrt2}(\ket H+\ket V)$}  & 
{$\frac{1}{\sqrt2}(\ket H-\ket V)$} & {$\cos\frac\pi8\ket H+\sin\frac\pi8\ket V$}  \\ \hline
$\tilde q=\<\Psi_{\rm in}|X_+\Psi_{\rm in}\>=$  &   0.509     & 0.509 & 0.948 & 0.033 & 0.843   \\
$q=\<\Psi_{\rm in}|B^{\pi/8}_+\Psi_{\rm in}\>=$  & 0.496 & 0.501 & 0.840 & 0.164 & 0.741 \\ \hline
$4|\tilde{q}-q|=\Delta_2(\tilde{\mathsf q},\mathsf q)^2=$ & 0.052 & 0.032   & 0.432 & 0.524 & 0.408 \\
 \end{tabular}
 \end{center}
These results are presented in Fig.~4 in the Letter.

\end{document}